\documentclass[twocolumn,pre,showpacs,preprintnumbers,amssymb,superscriptaddress]{revtex4-1}

\usepackage{epsfig, color}
\usepackage{amsthm, amsmath, amsfonts, ae, psfrag}
\usepackage{braket}
\usepackage{bbm}
\usepackage[english]{babel}

\usepackage{epstopdf} 

\newcommand{\bvec}[1]{{\bf\string#1}}

\usepackage[colorlinks,
pdfpagelabels,
pdfstartview = FitH,
bookmarksopen = true,
bookmarksnumbered = true,
linkcolor = blue,
plainpages = false,
hypertexnames = false,
citecolor = blue] {hyperref}

\begin{document}

\title{Modeling diffusion in colloidal suspensions by dynamical density functional theory using fundamental measure theory of hard spheres}

\author{Daniel~Stopper}

\email{daniel.stopper@uni-tuebingen.de}

\author{Kevin~Marolt}

\author{Roland~Roth}

\author{Hendrik~Hansen-Goos}

\affiliation{Institute for Theoretical Physics, University of T\"ubingen, Auf der Morgenstelle 14, 72076 T\"ubingen, Germany
}

\date{\today}

\begin{abstract}

We study the dynamics of colloidal suspensions of hard spheres that are subject to Brownian motion in the overdamped limit. We obtain the time evolution of the self and distinct parts of the van Hove function by means of dynamical density functional theory (DDFT). The free energy model for the hard sphere fluid that we use is the very accurate White Bear II version of Rosenfeld's fundamental measure theory. However, in order to remove interactions within the self part of the van Hove function a non-trivial modification has to be applied to the free energy functional. We compare our theoretical results with data that we obtain from dynamical Monte Carlo simulations and find that the latter are well described by our approach even for colloid packing fractions as large as 40\,\%.

\end{abstract}

\maketitle

\section{Introduction}

Colloidal suspensions provide a valuable test ground for theories that are based on classical statistical physics. Sized in the range of $100\text{ nm}$ to $10\text{ $\mu$m}$, colloids are small enough to undergo Brownian motion, thereby adequately exploring the phase space, while they are sufficiently large to be visualized using optical microscopy techniques. Most notably in recent years confocal microscopy has allowed one to track particle trajectories in large three dimensional samples. Interactions can be readily tuned by altering the surface chemistry of the colloidal particles as well as adjusting the ionic strength and index of refraction of the solvent. In particular, it is possible to obtain experimental conditions under which the colloids interact almost like hard spheres \cite{PuMe86}. A benefit of this experimental realization is that a variety of theoretical methods, such as integral equations \cite{Per57, Leb64, Kjel88}, mode coupling theory \cite{Goe09, Fu04}, and density functional theory \cite{Ev79,Ro10}, are capable of providing accurate results for the hard-sphere system, including phenomena such as the phase transition from the liquid to the crystal phase \cite{Che01}, the glass transition \cite{MeUn94}, fluid-fluid demixing in (non-additive) binary mixtures \cite{Dijk98}, only to mention a few. Obviously, all these examples pose challenges to statistical physics of equilibrium systems but even more so regarding particle dynamics; for instance the study of jamming in colloidal suspensions \cite{Trappe01} or structural relaxation near the glass transition \cite{Weeks00} calls for approaches that are based on non-equilibrium theories. In particular, the dynamics of hard-sphere like colloids has been extensively studied experimentally using different techniques such as dynamic light scattering and fluorescence recovery after photobleaching \cite{OtWi87,MeUn89,BlPeMaDh92,MeEA97}.

In the present work we devise a method for studying the dynamics of hard-sphere systems undergoing Brownian dynamics. We focus on the understanding of how to adapt an accurate equilibrium free energy model to this task, while hydrodynamic interactions, which are generally relevant in experimental systems, are not considered here as they can be included separately using established methods \cite{RoyallEA07,ReLoe09}. Our approach is based on the framework provided by dynamical density functional theory (DDFT) \cite{MarconiTarazona99, ArcherEvans04, Schmidt2007}. Building on a previous study by Hopkins {\em et al.} \cite{Schmidt2010}, which uses the rather simple Ramakrishnan-Yussouff (RY) free energy model for the inhomogeneous hard-sphere fluid \cite{RamYouss79}, we employ an accurate fundamental measure theory (FMT) type density functional for our free energy model \cite{Rf89, Ro10}. This makes our theory particularly reliable, especially in the case of dense suspensions.

The paper is structured as follows. In Section \ref{SecTheory} we give a brief summary of DDFT (\ref{SubSecDDFT}) followed by a presentation of different versions of FMT for the hard-sphere fluid (\ref{SubSecFMT}). For the present work we choose the so-called White Bear II version of FMT \cite{HaRo06} which is known to provide a very accurate account of the properties of inhomogeneous hard-sphere systems, both in the fluid \cite{Oe09} and crystal phase \cite{Oettel10}. Along the lines of the work by Hopkins {\em et al.} \cite{Schmidt2010} we then introduce the van Hove function (\ref{SubSecvanHove}) and the concept of dynamical test particle theory (\ref{SubSecDynamicTestTheory}) in order to provide the dynamical quantities that are at the center of our study, namely the self and distinct parts of the van Hove function. In the final part of the theory section we discuss how FMT can be used within DDFT (\ref{SubSecDDFTwithinFMT}). In order to insure that the theory is consistent with free diffusion in the low-density limit, interactions within the self component of the FMT need to be removed, which leads us to the partial-linearization approach.

In Section~\ref{SecImplementation} we discuss aspects of the numerical implementation of our DDFT followed by a brief account of dynamical Monte Carlo simulation \cite{SaMa10}, the method which we use in order to perform numerical experiments against which we validate our DDFT results.

Our results for the self and distinct parts of the van Hove function, as well as for the mean square displacement, from both the DDFT-FMT and the dynamical Monte Carlo simulations are presented in Section~\ref{SecResults}. The summary of our findings, along with our conclusion, can be found in the final Section~\ref{SecConclusion}.

\section{Theory} \label{SecTheory}

\subsection{Dynamical density functional theory (DDFT)} \label{SubSecDDFT}

The dynamical behavior of a system of $N$ classical (identical) colloids with positions $\mathbf{r}_i,~i = 1,..,N$ in the overdamped limit can be described by the following set of Langevin equations \cite{Langevin1908}:

\begin{equation} \label{langevinEq}
\Gamma^{-1} \frac{\text{d} \mathbf{r}_i}{\text{d} t} =  \mathbf{F}_{i} +  \mathbf{\eta}_i(t)~,~
\end{equation}
where $\Gamma^{-1}$ is a friction constant due to the motion of the Brownian particles through the solvent, $\mathbf{\eta}_i(t)$ is a stochastic force caused by random collisions of the solvent molecules with the colloids, fulfilling the condition $\braket{\mathbf{\eta}_i(t)} = 0$. Here $\braket{\cdot}$ denotes an average over initial conditions of the solvent. Moreover, $\mathbf{F}_i$ denotes the external force acting on particle $i$, which can be decomposed into the gradient of a inter-particle potential $U(\mathbf{r}_1,..,\mathbf{r}_N)$ and an arbitrary external potential $V_\text{ext}$. Assuming $U(\mathbf{r}_1,..,\mathbf{r}_N)$ to be a sum over a pair interaction potential, which depends only on the distance between two colloids $k$ and $j$, yields an expression for $\mathbf{F}_i$:  
\begin{equation} \label{ForceOnColloid}
\mathbf{F}_i = -\nabla_i \left(\frac{1}{2} \sum_{k = 1}^{N} \sum_{j \neq k}^{N} U(|\mathbf{r}_k - \mathbf{r}_j|) + \sum_{k = 1}^{N} V_\text{ext}(\mathbf{r}_k)  \right)
\end{equation}
Following Marconi and Tarazona \cite{MarconiTarazona99} the set of equations in Eq.~\eqref{langevinEq} can be rewritten as
\begin{align}\label{EqDynamicWithTwoBodyDensity}
\frac{\partial \rho(\mathbf{r}, t)}{\partial t} = & \,\,\nabla \left[ k_B T \nabla \rho(\mathbf{r}, t) + \rho(\mathbf{r}, t) \nabla V_\text{ext}(\mathbf{r})\right] \notag \\
 & + \nabla \left[ \int \text{d}^3 r'~ \rho^{(2)}(\mathbf{r}, \mathbf{r}', t) \nabla  U(|\mathbf{r} - \mathbf{r}'|) \right]~,~
\end{align}
where $\rho(\mathbf{r}, t)$ is the ``ensemble-averaged'' one-body density distribution 
\begin{equation}
\rho(\mathbf{r}, t) = \braket{\hat{\rho}(\mathbf{r}, t)} = \left\langle \sum_{i = 1}^{N} \delta(\mathbf{r}_i(t) - \mathbf{r}) \right\rangle~,~
\end{equation}
and $\rho^{(2)}(\mathbf{r}, \mathbf{r}', t) = \left\langle\hat{\rho}(\mathbf{r}, t) \hat{\rho}(\mathbf{r}', t) \right\rangle$ is the time-dependent two-body distribution function. Unfortunately, in general $\rho^{(2)}(\mathbf{r}, \mathbf{r}', t)$ is not known exactly. Therefore some approximations have to be made. One possibility is to make the assumption that the relation
\begin{equation}
k_B T \rho_0(\mathbf{r}) \nabla c^{(1)}(\mathbf{r}) = - \int \text{d}^3 r'~ \rho_0^{(2)}(\mathbf{r}, \mathbf{r}') \nabla  U(|\mathbf{r} - \mathbf{r}'|)~,~ \label{c1rho2Equil}
\end{equation}
where 
\begin{equation}
c^{(1)}(\mathbf{r})  =  - \beta \frac{\delta \mathcal{F}_\text{ex}[\rho(\mathbf{r})]}{\delta \rho(\mathbf{r})}~,~ \label{c1Equil}
\end{equation}
which holds for $\rho_0^{(2)}(\mathbf{r}, \mathbf{r}')$ in thermodynamic equilibrium, can be used as an approximation also for non-equilibrium systems \cite{MarconiTarazona99, ArcherEvans04}. Here $\rho_0(\mathbf{r})$ denotes the one-body density distribution in equilibrium, $c^{(1)}(\mathbf{r})$ is referred to as the direct one-body correlation function which is given by the first functional derivative of the excess free energy functional $\mathcal{F}_\text{ex}[\rho]$, and $\beta = 1/k_B T$ is the inverse temperature.

Substituting Eqs.~\eqref{c1rho2Equil} and \eqref{c1Equil} into Eq.~\eqref{EqDynamicWithTwoBodyDensity} yields the key equation of dynamical density functional theory
\begin{equation} \label{EqDDFT}
\frac{\partial \rho(\mathbf{r}, t)}{\partial t} = \Gamma \nabla \left[ \rho(\mathbf{r}, t) \nabla \frac{\delta \mathcal{F}_\mathcal{H}[\rho(\mathbf{r}, t)]}{\delta  \rho(\mathbf{r}, t)} \right]~,~
\end{equation}
with the so-called Helmholtz free energy functional $\mathcal{F}_\mathcal{H}$ which has the form 
\begin{align} \label{EqHelmholtzFreeEnergy}
\mathcal{F}_\mathcal{H}[\rho(\mathbf{r}, t)]
 = & \,\, k_B T\int \text{d}^3 r~ \rho(\mathbf{r}, t)  \left( \ln \left( \lambda^3 \rho(\mathbf{r}, t) \right) - 1 \right) \notag \\ 
 &+ \mathcal{F}_\text{ex}[\rho(\mathbf{r}, t)] + \int \text{d}^3 r~\rho(\mathbf{r}, t) V_\text{ext}(\mathbf{r}, t)\, .
\end{align}
The first term in Eq.~\eqref{EqHelmholtzFreeEnergy} is the ideal-gas contribution, $\lambda$ is the thermal wavelength. Note that in this derivation of DDFT the mobility $\Gamma$ is assumed to be a constant in space and time and is linked to the Stokes-Einstein diffusion coefficient by $D = k_B T \Gamma$ \cite{Einstein1905}. 

The multi-component generalization of Eq.~\eqref{EqDDFT} to a system consisting of $\nu$ species of colloids with radii $R_i, i = 1,..,\nu$ is given by \cite{Archer05,RoRaAr09}
\begin{equation} \label{eq_ddft_comp}
\frac{\partial \rho_i(\mathbf{r}, t)}{\partial t} = \Gamma_i \nabla \left[ \rho_i(\mathbf{r}, t) \nabla \frac{\delta \mathcal{F}_\mathcal{H}[\{ \rho_j\}]}{\delta  \rho_i(\mathbf{r}, t)} \right]~,~
\end{equation}
where the Helmholtz free energy functional now takes the form
\begin{align}
 \mathcal{F}_\mathcal{H}[\{ \rho_i\}] = & \,\, k_B T \sum_{i = 1}^{\nu} \int \text{d}^3 r~ \rho_i(\mathbf{r}, t) \left[ \ln \left( \lambda_i^3 \rho_i(\mathbf{r}, t) \right) - 1 \right] \notag \\
 &+ \mathcal{F}_\text{ex}[\{ \rho_i \}] + \sum_{i = 1}^{\nu} \int \text{d}^3 r~ \rho_i(\mathbf{r}, t) V^i_\text{ext}(\mathbf{r}).
\end{align}

Note that standard DDFT does not include memory effects. In order to describe non-Markovian dynamics, Brader and Schmidt have recently put forward the framework of power functional theory \cite{BrSm13,BrSm14,BrSm15}. However, it is presently not clear how accurate approximations to the excess dissipation functional underlying the approach should be constructed.

\subsection{Fundamental measure theory (FMT)} \label{SubSecFMT}
As in equilibrium density functional theory (DFT) \cite{Ev79} an important task in DDFT is to use a reliable approximation to the (generally) unknown excess free energy functional $\mathcal{F}_\text{ex}$.
Here we use the accurate White Bear II (WBII) functional which is based on fundamental measure theory (FMT) for hard-sphere mixtures \cite{Rf89, HaRo06}. In what follows we recall the basics of FMT introduced by Rosenfeld in 1989 and its extensions - for a more detailed account see Refs.~\onlinecite{Rf89, Rf93, Rf96, Rf97, Tarazona2000, RoEvLaKa02, HaRo06}.

\subsubsection*{The Rosenfeld functional}

At the center of Rosenfeld's FMT is the observation that in a dilute hard-sphere mixture the excess free energy functional can be written as
\begin{equation}\label{EqLowDensityF_ex}
 \beta \mathcal{F}_\text{ex} 
 =\int \text{d}^3 r \left[ n_0(\mathbf{r})n_3(\mathbf{r}) + n_1(\mathbf{r})n_2(\mathbf{r}) - \mathbf{n}_1(\mathbf{r})\mathbf{n}_2(\mathbf{r}) \right]~,~
\end{equation}
a result which was obtained by expressing the Mayer-$f$-function of the hard-sphere fluid in terms of geometrical properties of the overlap of two spheres \cite{Rf88,Rf89}. Here the $n_{\alpha}$ denote weighted densities, which are calculated as
\begin{equation} \label{EqWeightedDens}
 n_\alpha(\mathbf{r}) = \sum_{i = 1}^{\nu}\int \text{d}^3r'~\rho_i(\mathbf{r}') \omega_\alpha^i(\mathbf{r} - \mathbf{r}')
\end{equation}
using the weight functions
\begin{align} \label{EqDefWeightFunctions}
 \omega_3^i (\mathbf{r}) &= \Theta(R_i - |\mathbf{r}|)& \omega_2^i(\mathbf{r}) &= \delta(R_i - |\mathbf{r}|) \\
 \omega_1^i (\mathbf{r}) &= \frac{\omega_2^i}{4\pi R_i}&  \omega_0^i(\mathbf{r}) &= \frac{\omega_2^i}{4\pi R_i^2} \\
 \boldsymbol{\omega}_2^i(\mathbf{r}) &= \frac{\mathbf{r}}{r} \omega_2^i&  \boldsymbol{\omega}_1^i(\mathbf{r}) &= \frac{\boldsymbol{\omega}_2^i}{4\pi R_i}~,~
\end{align}
where index $\alpha$ labels four scalar and two vectorial weighted densities, $\Theta(\cdot)$ denotes the Heaviside function, and $\delta(\cdot)$ is the Dirac delta function. Note that weighted densities akin to the $n_\alpha(\mathbf{r})$ also occur in the exact one-dimensional functional for hard-rod mixtures \cite{Vanderlick89}. 

In order to derive an excess free energy functional at larger densities, Rosenfeld used the following ansatz:
 \begin{equation} \label{EqF_exAnsatz}
 \beta \mathcal{F}_\text{ex}[\{\rho_i\}] =\int \text{d}^3r~\Phi(\{n_\alpha(\mathbf{r})\})~,~
 \end{equation}
 which can also be motivated by the exact result in one dimension, where the excess free energy density $\Phi$ is a {\em function} of certain weighted densities.
Equation \eqref{EqF_exAnsatz} yields the following form of the direct one-body correlation function $c^{(1)}(\mathbf{r})$ of species $i$ (see Eq.~\eqref{c1Equil} for a definition)
   \begin{equation} \label{Eq_c^1withRfansatz}
   c^{(1)}_i(\mathbf{r}) = - \sum_{\alpha} \int \text{d}^3r'~ \frac{\partial \Phi(\{n_\alpha\})}{\partial n_\alpha} \frac{\delta n_\alpha (\mathbf{r}')}{\delta \rho_i(\mathbf{r})}.
\end{equation} 

In order to determine the function $\Phi$, Rosenfeld used dimensional analysis, the condition that Eq.~\eqref{EqF_exAnsatz} has to recover the low-density expansion, Eq.~\eqref{EqLowDensityF_ex}, and an exact relation from scaled-particle theory, which reads
 \begin{equation} \label{EqSPTCondition}
 	\lim_{R_i \rightarrow \infty} \frac{\beta \mu_\text{ex}^i}{\frac{4\pi}{3} R_i^3} =\beta p~,~
 \end{equation} 
relating the work of reversibly introducing a large sphere into the fluid (i.e. the excess chemical potential $\mu_\text{ex}^i$) to the pressure $p$. The final result obtained by Rosenfeld is \cite{Rf89}
  \begin{align} \label{EqF_exRosenfeld}
  \Phi &= -n_0 \ln(1 - n_3) + \frac{n_1 n_2 - \mathbf{n}_1 \cdot \mathbf{n}_2}{1-n_3} + \frac{n_2^3 - 3n_2 \mathbf{n}_2 \cdot  \mathbf{n}_2}{24 \pi (1-n_3)^2} \notag \\ &\equiv \Phi_1 + \Phi_2 + \Phi_3.
  \end{align} 

\subsubsection*{$q_3$ correction and tensorial functional}

While being very successful in describing many aspects of the inhomogeneous hard-sphere fluid, including mixtures, it turns out that Rosenfeld's free energy density in its original form is not able to describe a hard-sphere crystal \cite{Rf89, Rf93}. A negative divergence occurs in the final term $\Phi_3$ for strongly peaked density profiles. In order to regularize $\Phi_3$, Rosenfeld {\it et al.} suggested to modify the term $\Phi_3$ as follows \cite{Rf96, Rf97}:
   \begin{equation} \label{EqQ3Corretion}
   \tilde{\Phi}_3 = \frac{1}{24 \pi (1 - n_3)^2} \left( n_2 - \frac{\mathbf{n}_2 \cdot \mathbf{n}_2}{n_2} \right)^3~,~
   \end{equation} 
which is referred to as the $q_3$ correction.
  An alternative approach that regularizes $\Phi_3$ was given by Tarazona who introduced an additional {\em tensorial} weight function, and thus a tensorial weighted density \cite{Tarazona2000}. These are given by (we follow the notation of Ref.~\onlinecite{Schmidt2000})
  \begin{equation} \label{EqTensWeightFunc}
  \omega_{m_2}(\mathbf{r}) = \left( \frac{\mathbf{r} \otimes \mathbf{r}}{r^2} - \frac{1}{3} \mathbbm{1} \right) \omega_2(\mathbf{r})~,~
  \end{equation}
  and
  \begin{equation} \label{EqTensWeightDens}
  n_{m_2}(\mathbf{r}) = \int \text{d}^3 r ~\rho(\mathbf{r}) \omega_{m_2}(\mathbf{r} - \mathbf{r}').
  \end{equation}
Here $\mathbbm{1}$ denotes the $3 \times 3$ unity matrix and $\mathbf{r}\otimes\mathbf{r}$ represents the dyadic product of two vectors. $\Phi_3$ is replaced by a new term $  \Phi_3^\text{\tiny T}$ containing $n_{m_2}$:
  \begin{equation}
  \Phi_3^\text{\tiny T} = \Phi_3 + \frac{ \frac{9}{2} \left( \mathbf{n}_2 n_{m_2} \mathbf{n}_2 - \text{Tr} \left( n_{m_2}^3 \right) \right)  }{24 \pi (1 - n_3)^2}~,~
  \end{equation}
where Tr$(\cdot)$ denotes the trace of a matrix. 

Both the $q_3$ correction and the tensorial modification lead to functionals that give decent descriptions of the hard-sphere crystal, while leaving the already excellent properties of the functional regarding the descriptions of the fluid phase virtually unaffected.
  
  \subsubsection*{White Bear versions of FMT}
  
  By applying the FMT which we have outlined above to the bulk fluid, thermodynamic properties such as the pressure can be derived. It turns out that using Rosenfeld's free energy density results in the generalized Percus-Yevick (PY) compressibility equation of state for the hard-sphere mixture.
  
  However, it is possible to use the more accurate Mansoori-Carnahan-Starling-Leland (MCSL) equation of state \cite{Mansoori1971}, which is a $\nu$-component generalization of the well-known Carnahan-Starling (CS) equation of state \cite{CaSt69}, as an input to derive a new excess free energy density $\Phi^{\text{\tiny WB}}$ - the White Bear (WB) version of FMT \cite{RoEvLaKa02}. The WB functional performs better in describing density profiles of hard sphere mixtures, especially at high bulk densities close to freezing transition \cite{RoEvLaKa02}. However, it is found that the WB functional does not recover Eq.~\eqref{EqSPTCondition}, meaning that the partial derivative of $\Phi^{\text{\tiny WB}}$ with respect to $n_3$ does not give rise to the equation of state originally used for the derivation of $\Phi^{\text{\tiny WB}}$. Obviously, this inconsistency with scaled-particle theory has to be expected because, as discussed above, using Eq.~\eqref{EqSPTCondition} in order to determine $\Phi$ precisely leads one to the less accurate PY equation of state.   
  
In order to minimize this inconsistency, a new generalization of the Carnahan-Starling equation of state has been put forward \cite{HaRo06eos}. Based on this new equation of state it is possible to derive the following functional \cite{HaRo06}:
  \begin{align}
  \Phi^{\text{\tiny WBII}} &= -n_0 \ln(1 - n_3) + \left( n_1 n_2 - \mathbf{n}_1 \cdot \mathbf{n}_2 \right) \frac{1 + \frac{1}{3}\phi_2(n_3)}{1 - n_3} \notag \\
  &+ \left( n_2^3 - 3n_2 \mathbf{n}_2 \cdot \mathbf{n}_2 \right) \frac{1 - \frac{1}{3}\phi_3(n_3)}{24\pi (1 - n_3)^2}~,~
  \end{align}
  in which the functions $\phi_2$ and $\phi_3$ are given by
  \begin{align}
  \phi_2(n_3) &= \frac{1}{n_3} \left(  2 n_3 - n_3^2 + 2(1 - n_3)\ln(1 - n_3) \right) \notag ~,~ \\
  \phi_3(n_3) &= \frac{1}{n_3^2}\left( 2 n_3 - 3n_3^2 + 2n_3^3 + 2(1 - n_3)^2\ln(1 - n_3) \right).
  \end{align}
This functional, the White Bear version mark II, is consistent with the scaled-particle relation, i.e. $\beta p_\text{\tiny CS} = \partial \Phi/\partial n_3$ in the case of the one-component fluid. Since the derivations of the White Bear versions of FMT start from the same ansatz as Rosenfeld's FMT (mainly they differ in the choice of the equation of state) it is obvious that they face the same problems when describing a hard sphere crystal. However, one can also apply the empirical $q_3$ correction or the tensorial approach due to Tarazona to $\Phi_3$. In particular the tensorial WBII functional has been demonstrated to provide an excellent description of hard sphere crystals \cite{Oettel10}.
  
\subsection{van Hove function} \label{SubSecvanHove}

A convenient quantity for the description of diffusion is provided by the van Hove function \cite{vanHove54}, which we review in the following. To this end we note that the probability of finding a particle located at position $\mathbf{r} + \mathbf{r}'$ at time $t > 0$  given that another particle was located at $\mathbf{r}'$ at time $t = 0$ can be written as \cite{HaMcDo13}
\begin{equation}
\mathcal{G}(\mathbf{r}, \mathbf{r}', t) = \frac{1}{N} \left\langle \hat{\rho}(\mathbf{r} + \mathbf{r}', t) \hat{\rho}(\mathbf{r}', 0) \right\rangle.
\end{equation}
Eliminating the choice of origin by integrating over $\mathbf{r}'$ yields the van Hove function \cite{HaMcDo13, vanHove54}
\begin{align}  \label{EqvanHove}
\mathcal{G}(\mathbf{r}, t) &= \frac{1}{N} \left\langle  \int \text{d}^3 r'~ \hat{ \rho}(\mathbf{r}' + \mathbf{r}, t)\hat{\rho}(\mathbf{r}', 0) \right\rangle \\
&= \frac{1}{\rho_b} \left\langle \hat{\rho}(\mathbf{r}, t) \hat{\rho}(\mathbf{0}, 0) \right\rangle \, , \label{EqvanHoveUniform}
\end{align}
where $\rho_b=\frac{N}{V}$ is the particle number density of the bulk fluid. In this representation $\mathcal{G}(\mathbf{r}, t)$ is referred to as a dynamic density-density auto-correlation function, where Eq. \eqref{EqvanHoveUniform} holds only in the case of a uniform fluid. Thus, the van Hove function gives the probability of finding an arbitrary particle located at position $\mathbf{r}$ at time $t$ provided a particle has been at the origin at time $t = 0$. Using the properties of the Dirac delta function $\delta(\cdot)$, it is easy to derive the more common representation of the van Hove function: 
\begin{equation} \label{EqvanHoveCommon}
\mathcal{G}(\mathbf{r}, t) = \frac{1}{N} \left\langle  \sum_{i = 1}^{N} \sum_{j = 1}^{N} \delta\left(\mathbf{r} + \mathbf{r}_j(0) - \mathbf{r}_i(t)\right) \right\rangle.
\end{equation}
Equation~\eqref{EqvanHoveCommon} naturally splits into two parts by discriminating between the cases $i \neq j$ and $i = j$:
\begin{align}
\mathcal{G}(\mathbf{r}, t) &= \frac{1}{N} \left\langle  \sum_{i = 1}^{N} \delta\left(\mathbf{r} + \mathbf{r}_i(0) - \mathbf{r}_i(t)\right) \right\rangle \notag \\ 
&+ \frac{1}{N} \left\langle  \sum_{i  = 1}^{N} \sum_{j \neq  i}^{N} \delta\left(\mathbf{r} + \mathbf{r}_j(0) - \mathbf{r}_j(t)\right) \right\rangle \notag \\
&\equiv \mathcal{G}_s (\mathbf{r}, t) + \mathcal{G}_d(\mathbf{r}, t).
\end{align}
The terms $\mathcal{G}_s(\mathbf{r}, t)$ and $\mathcal{G}_d(\mathbf{r}, t)$  are referred to as the self ($s$) and distinct ($d$) parts of the van Hove function. The self part characterizes the behavior of the particle initially located at the origin, whereas the distinct part describes the average motion of the remaining particles. We expect the number of particles to be a conserved quantity in time which can easily be shown by considering the volume integrals of  $\mathcal{G}_s(\mathbf{r}, t)$ and $\mathcal{G}_d(\mathbf{r}, t)$. 

In what follows we are interested in an homogeneous and isotropic fluid, thus the dependency of $\mathcal{G}(r, t)$ is only on the distance to the origin $r = |\mathbf{r}|$. At time $t = 0$ we find
\begin{equation}
\mathcal{G}(r, 0) = \delta(\mathbf{r}) + \rho_b g(r) = \mathcal{G}_s(r, 0) + \mathcal{G}_d(r, 0)~,~
\end{equation}
where $g(r)$ denotes the radial distribution function. It gives the probability of finding a particle at distance $r$ to a reference point $\mathbf{r}_i$, given that another particle  is located at position $\mathbf{r}_i$ (for more detailed account see e.g.\ Ref.~\onlinecite{HaMcDo13}). The asymptotic behavior of the self and distinct part in the thermodynamic limit is given by \cite{HaMcDo13} 
\begin{align}
\lim_{r \rightarrow \infty} \mathcal{G}_s(r, t) &= \lim_{t \rightarrow \infty} \mathcal{G}_s(r, t) = 0~,~ \\
\lim_{r \rightarrow \infty} \mathcal{G}_d(r, t) &= \lim_{t \rightarrow \infty} \mathcal{G}_d(r, t) = \rho_b \label{EqvanHoveDistinctLimit}.
\end{align}

As long as the system is in the fluid phase, i.e.\ densities are sufficiently low so that the diffusion process is not disturbed due to trapping effects, it can be shown that in the long-time limit the self-part is of Gaussian shape
\begin{equation} \label{EqAproxLowDensG_s}
\mathcal{G}_s(r, t) = \left( \frac{1}{4\pi D_lt} \right)^{\frac{3}{2}} \exp \left( - \frac{r^2}{4D_l t} \right).
\end{equation}
with a mean-square displacement given by
\begin{equation} \label{EqMSQD}
\left\langle r^2 \right\rangle(t) = 4 \pi \int_{0}^{\infty} \text{d}r~r^4 \mathcal{G}_s(r, t) = 6D_lt~,~
\end{equation}
where $D_l$ is referred to as self (or long time) diffusion coefficient and is not equal to the Einstein diffusion coefficient $D$ \cite{Einstein1905}, which describes the diffusion of a single particle within a solvent.

Only for low colloid densities $\rho_b \rightarrow 0$, where interactions between the colloids are negligible, can we identify $D_l$ with $D$. In this case we may consider our system as an ideal gas, therefore the mean square displacement is $\left\langle r^2 \right\rangle(t) = 6 D t$. The Einstein diffusion coefficient $D$ is connected to the Brownian time via $\tau_B = \sigma^2/D$ where $\sigma$ is the diameter of the colloid \cite{HaMcDo13} and $\tau_B$ can be understood as the time that it takes a colloid to diffuse a distance comparable to its size. Hence the Brownian time $\tau_B$ is the time scale which is relevant for the dynamics of a colloidal suspension.

\subsection{Dynamical test particle limit} \label{SubSecDynamicTestTheory}

In order to calculate the self and distinct parts of the van Hove function, we use DDFT together with a dynamical extension of Percus'
test particle limit \cite{Percus62, Schmidt2007, Schmidt2010}. Percus showed that if one considers a test particle in equilibrium, then the one-body density distribution $\rho(r)$ of the surrounding particles is closely related to the radial distribution function $g(r)$ if and only if one sets the external potential acting on the fluid equal to the pair interaction potential of the particles:
\begin{equation} \label{EqPercusTestLimit}
\rho(r) = \rho_b g(r).
\end{equation}

 The extension of Percus' approach to dynamic processes was first tackled in Ref.~\onlinecite{Schmidt2007}. Following Ref.~\onlinecite{Schmidt2007, Schmidt2010} we treat our system as a binary mixture of species $s$ (self) and $d$ (distinct). Moreover, we assume that species $s$ consists of only one test particle whereas species $d$ consists of the remaining $N-1$ particles. Now consider a particle of species $s$ located at the origin at time $t = 0$. Hence, its density distribution is given by $\rho_s(r,t = 0) = \delta(\mathbf{r})$. According to Percus, the density distribution of species $d$ then reads $\rho_d(r, t = 0) = \rho_b g(r)$. For times $t > 0$ we assume the coordinate system to be fixed in space and  we are interested in the behavior of the binary mixture, i.e. we are interested in the density distributions $\rho_s(r, t)$ and $\rho_d(r, t)$. Based on the definitions of the self and distinct parts of the van Hove function $\mathcal{G}(r, t)$ we can identify 
\begin{align}
\rho_s(r,t) &\equiv \mathcal{G}_s(r,t) ~,~ \\
\rho_d(r,t) &\equiv \mathcal{G}_d(r,t)~,~
\end{align}
for all times $t \geq 0$. The initial condition of species $d$ at time $t = 0$ can be determined from equilibrium DFT by minimizing the functional of the grand potential $\Omega[\rho]$.

In principal, we are now able to calculate the van Hove function by means of DDFT, using Eq. \eqref{eq_ddft_comp} where $i = s, d$ and a FMT based excess free energy functional. Since the self part describes the behavior of a single particle, we first have to ensure that hard-sphere interactions within the self part are removed. In the next section, we therefore introduce two different modifications that we have applied to the excess free energy functional $\mathcal{F}_\text{ex}$ in order to remove interactions within the self part. Finally, it is worthwhile to note that treating the system as a binary mixture of self and distinct particles does allow one to study more general situations where the self and distinct particles might differ in size and/or shape. In this regard we note that FMT free energy formulations are available for fluid mixtures of arbitrarily shaped hard particles \cite{Ros94, HaMe09}.

\subsection{DDFT formulation using FMT} \label{SubSecDDFTwithinFMT}

The first approach that we use in this paper in order to remove interactions within the self part of the functional is based on the density functional for a colloid-polymer mixture \cite{Schmidt2000}. The interactions between colloids ($c$) and polymers ($p$) underlying this functional are $U_{cc}=U_{cp}=U_{pc}=U_{\text{hs}}$, where $U_{\text{hs}}$ is the usual hard-sphere interaction potential and $U_{pp} = 0$. The derivation of the functional is based on the zero-dimensional limit, i.e.\ a cavity which can hold at most one colloid but can hold an arbitrary number of polymers if no colloid is present. In particular, it has been shown that such a colloid-ideal polymer functional can be derived by means of linearizing the Rosenfeld functional (and hence any FMT based functional) with respect to the polymer component. An obvious mapping to our problem at hand consists in mapping the colloid species to the distinct part ($c\leftrightarrow d$) and the polymer species to the self part ($p\leftrightarrow s$). This insures that the self part (like the polymers) does not experience interactions while all the other  interactions are of the hard-sphere type as they should. Thus, we modify the excess free energy density as follows:
\begin{equation}
 \Phi \left( \{ n_\alpha^s, n_\alpha^d \}, t \right) \rightarrow \tilde{\Phi}\left( \{ n_\alpha^s, n_\alpha^d \}, t \right)~,~
\end{equation}
 in which $\tilde{\Phi}$ is given by
\begin{equation} \label{EqPhi_linearized}
  \tilde{\Phi}\left( \{ n_\alpha^s, n_\alpha^d \}, t \right)  = \left.\Phi\left( \{ n_\alpha^s, n_\alpha^d \}, t\right)\right|_{n_\alpha^s=0} + \sum_\alpha \left.\frac{\partial \Phi}{\partial n_\alpha^s} \right|_{n_\alpha^s=0} \, n_\alpha^s.
\end{equation}
Equation~\eqref{Eq_c^1withRfansatz} implies that the direct pair correlation functions together with Eq. \eqref{EqPhi_linearized}  take the following form:
 \begin{align} 
 c^{(1)}_s(\mathbf{r}, t) &= - \sum_{\alpha} \int \text{d}^3 r'~ \left. \frac{\partial \Phi(\mathbf{r}', t)}{\partial n_\alpha^s} \right|_{n_\alpha^s=0} \omega_\alpha^s(\mathbf{r}' - \mathbf{r}) ~,~ \label{c1_s_linearized}\\
  c^{(1)}_d(\mathbf{r}, t) &= - \sum_{\alpha} \int \text{d}^3 r'~  \frac{\partial \tilde{\Phi}(\mathbf{r}', t)}{\partial n_\alpha^d}  \omega_\alpha^d(\mathbf{r}' - \mathbf{r}) \label{c1_d_linearized}.
 \end{align}
 We see that $c^{(1)}_s(\mathbf{r}, t)$ depends only on the weighted densities $n_\alpha^d$ of species $d$ which is due to the fact that the test particle interacts only with its surrounding particles whereas $c^{(1)}_d(\mathbf{r}, t)$ contains the information about distinct-distinct as well as self-distinct interactions. 
 By substituting these results into Eq. \eqref{eq_ddft_comp} or equivalently into
 \begin{align} \label{EqDDFTAlternative}
 \frac{\partial \rho_i(\mathbf{r}, t)}{\partial t} = D_i \nabla \left[ \nabla \rho_i(\mathbf{r}, t) -  \rho_i(\mathbf{r}, t) \nabla c^{(1)}_i(\mathbf{r}, t) \right ] 
 \end{align}
one can start to perform numerical calculations in order to determine the time evolution of the van Hove function.

However, note that while the mapping of our system onto the colloid-polymer mixture does indeed give us the correct interactions between the species, there is an important difference that should be kept in mind. The functional of Schmidt {\it et al.} \cite{Schmidt2000} has been derived within the grand canonical ensemble, i.e.\ if we assume a density of the self particles (or, equivalently of the polymers) of $\eta_s=1$ in the zero dimensional cavity considered for its derivation, we assume that there is one self particle {\em on average}. Configurations without a self-particle or with more than one self particle do have a significant statistical weight. This obviously differs from the situation that we are attempting to model: there is {\em always exactly one} self particle present in the system. In this sense the self particle should be considered in the canonical ensemble while the distinct particles are well represented in the grand canonical ensemble. As a result, we might expect an incorrect dynamical behavior of the distinct part $\mathcal{G}_d(r, t)$ because due to the significant weight of configurations with no self particle present the space initially occupied by the self particle can easily be ``invaded'' by the distinct species. Technically, this is reflected in the fact that due to the linearization of the functional about the density of the self species the constraint on the local packing fraction $\eta_s+\eta_d<1$, which is encoded in the singularity of the functional at $n_3=n_3^s+n_3^d=1$ before linearization, is lost.

 \begin{figure}[t!] 
 \centering
 \includegraphics[width=9cm]{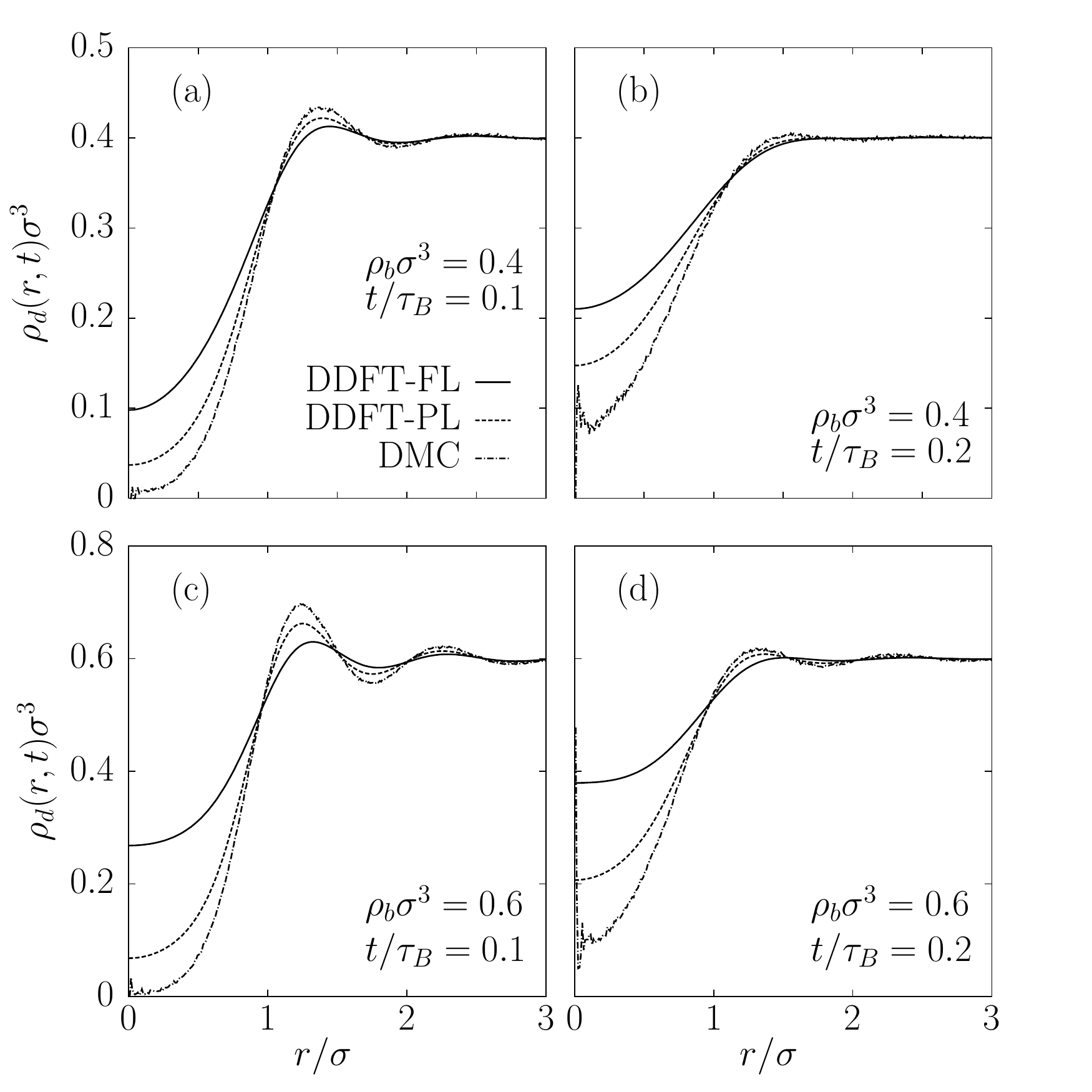}
 \caption{The distinct density profiles $\rho_d(r,t)$ for a hard-sphere fluid obtained by means of dynamical test particle theory and a full linerization (FL) vs.\ partial linearization (PL) of the tensorial White Bear II functional with respect to the self component, see text for details. The dashed-dotted lines show the profiles obtained by dynamical Monte-Carlo simulations. (a) and (b) display the results for a bulk density $\rho_b\sigma^3 = 0.4$, (c) and (d) for a density $\rho_b\sigma^3 = 0.6$. The times are  $t/\tau_B = 0.1$ in (a) and (c) and $t/\tau_B = 0.2$ in (b) and (d).
}
 \label{Fig_fullVsPartLin}
 \end{figure}

This is indeed what we observe in our numerical calculations, see Sec.~\ref{SubSecDiscretization} regarding the implementation. For illustration, in Fig. \ref{Fig_fullVsPartLin} we show the typical dynamic behavior of $\rho_d(r,t)$ that we obtain using Eq. \eqref{c1_d_linearized} (with the tensorial WBII functional) for bulk densities $\rho_b = 0.4$ and $\rho_b = 0.6$ at times $t/\tau_B = 0.1~\text{and}~ 0.2$. The behavior is qualitatively correct since we expect $\rho_d(r,t)$ to converge to $\rho_b$, see Eq. \eqref{EqvanHoveDistinctLimit}, but quantitatively it is not very satisfying. In comparison with data obtained from our dynamic Monte-Carlo simulations, see Sec.~\ref{SubSecDMC} for details, one can conclude that the structure of the density profiles obtained from dynamic test particle theory vanishes too rapidly. This effect is found by us at all bulk densities becoming more pronounced at higher densities. In particular, the profiles for $\rho_d(r,t)$ predict rather large densities close to the test particle ($r \lesssim \sigma$) already at times $t\lesssim \tau_B$ for which the test particle should not have diffused much from its original starting point. As mentioned earlier, this effect is probably a consequence of the fact that the statistical weight of configurations without a particle of species $s$ does not vanish in the treatment within the grand canonical ensemble.

In order to mitigate the shortcomings arising from the use of the fully linearized functional, we put forward a second, empirical ansatz for the direct one-body correlation functions:
    \begin{align} 
    c^{(1)}_s(\mathbf{r}, t) &= - \sum_{\alpha} \int \text{d}^3 r'~ \left. \frac{\partial \Phi(\mathbf{r}', t)}{\partial n_\alpha^s} \right|_{n_\alpha^s=0} \omega_\alpha^s(\mathbf{r}' - \mathbf{r})~,~  \label{Eqc1_sHalfLinearized} \\
     c^{(1)}_d(\mathbf{r}, t) &= - \sum_{\alpha} \int \text{d}^3 r'~  \frac{\partial \Phi(\mathbf{r}', t)}{\partial n_\alpha^d}  \omega_\alpha^d(\mathbf{r}' - \mathbf{r}) \label{Eqc1_dNotLinearized}.
    \end{align}
Here Eq.~\eqref{Eqc1_sHalfLinearized} remains unchanged relative to Eq.~\eqref{c1_s_linearized} which is in line with the argument that from the perspective of the self particle (or, equivalently, test particle) interactions occur only with the surrounding distinct particles. In order to avoid the violation of the local packing constraint $\eta_d+\eta_s<1$ we suggest to calculate $c^{(1)}_d(\mathbf{r}, t)$ from the full functional $\Phi(\{n_\alpha^d, n_\alpha^s\}, t)$, see Eq. \eqref{Eqc1_dNotLinearized}, rather than using the linearized functional $\tilde{\Phi}(\mathbf{r}', t)$ as in Eq.~\eqref{c1_d_linearized}. Henceforth, we shall refer to this procedure as the partial-linearization approach.

In Fig.~\ref{Fig_fullVsPartLin} we display the distinct part $\rho_d(r,t)$ for bulk densities $\rho_b\sigma^3 = 0.4$ and $\rho_b\sigma^3 = 0.6$ at times $t/\tau_B = 0.1$ and $t/\tau_B = 0.2$. We see that the results from the partial-linearization approach for the White Bear II functional agree much better with the simulations than those obtained using a full linearization of the functional. In particular, the distinct density $\rho_d(r,t)$ does not increase as strongly for $r\lesssim\sigma$, reflecting the implementation of the packing constraint, while the profiles at larger distances $r > \sigma$ maintain their structure for longer times, compared to the result from the full linearization. We find that using the partial-linearization approach leads to significant improvement at all densities, especially for $\rho_b\sigma^3 \gtrsim 0.4$. Nevertheless, the density profiles still lose their structure somewhat too fast compared to the simulation results. We will discuss possible reasons for this behavior in Sec. \ref{SecConclusion}.

\section{Implementation} \label{SecImplementation}

\subsection{Discretization of the DDFT} \label{SubSecDiscretization}

In what follows all results are obtained using the very accurate tensorial White Bear II excess free energy functional within the partial-linearization route. The initial condition of the self part $\rho_s(r,0) = \delta(\bvec{r})$ is implemented numerically by a strongly peaked Gaussian distribution,
\begin{equation}
\rho_s(r, 0) = \left( \frac{\alpha}{\pi} \right)^{\frac{3}{2}} \exp( - \alpha r^2)~,~
\end{equation} 
where we use $\alpha = 10^3$.  The initial profile of the distinct part $\rho_d(r, 0)$ is obtained by means of equilibrium DFT with a spherical external potential corresponding to the ``tagged'' test particle in the origin. Further discussions of numerical details in order to minimize the functional of the grand canonical potential can be found e.g. in Ref.~\onlinecite{Ro10}. 

In order to determine the dynamic behavior of the self and distinct parts of the van Hove function we integrate Eq.~\eqref{EqDDFTAlternative} forward in time using the Euler-forward algorithm with time steps of $\Delta t = 10^{-5} \tau_B$. We consider in this paper particles of the same radius, $R_s = R_d = R$. In case of spherical particles the Brownian time is given by $\tau_B = \sigma^2/D$ \cite{HaMcDo13} with $\sigma = 2R$ and the Einstein diffusion coefficient $D = \Gamma k_B T$. The total integration time is in general $t_\text{max} = \tau_B$. The partial derivatives $\partial_r$ as well as $\partial^2_r$ which occur in Eq. \eqref{EqDDFTAlternative} are calculated numerically. Moreover, the spatial integrations in Eqs.~\eqref{c1_s_linearized} and \eqref{Eqc1_dNotLinearized} are performed using the trapezoidal rule. The spatial resolution is $\delta x = 10^{-2}\sigma$.

The DDFT results for the mean square displacement shown in Fig.~\ref{Fig_width1.0} were obtained from a numerical integration of the density profiles $\rho_s(r,t)\equiv \mathcal{G}_s(r,t)$ according to Eq.~\eqref{EqMSQD}.

\subsection{Dynamic Monte-Carlo simulation (DMC)} \label{SubSecDMC}

The dynamic Monte-Carlo method (DMC) is based on the property of standard Monte-Carlo (MC) simulations to provide a representation of Brownian dynamics in the limit of small displacements $\delta l$ at each trial move. This can be understood from the observation that via the mean square displacement of a free particle the displacement $\delta l$ is associated with a time $\delta t$ thereby linking MC time to the Brownian timescale $\tau_{B}$ of the system. Convergence of MC simulations to Brownian dynamics can be significantly enhanced by rescaling MC time with the acceptance rate (see Ref.~\onlinecite{SaMa10} for details). We use this mapping between rescaled MC time and Brownian time in order to obtain the DMC results presented in this work.

In order to model the systems with densities $\rho_b \sigma^3 = 0.2 \ldots 0.8$ we use a box of dimension $10 \sigma \times 10 \sigma \times 10 \sigma$ with a number of particles ranging from $N = 200$ to $N = 800$. The maximum displacement $\delta l$ of a particle upon a MC step is chosen such that the associated time $\delta t$ ranges from $10^{-4}\tau_{B}$ for the low-density systems to $10^{-5}\tau_{B}$ for the highest density considered here ($\rho_b\sigma^3 = 0.8$). We have verified that the values for $\delta l$ employed here are sufficiently small in order to guarantee a convergence of the DMC simulations to Brownian dynamics. The number of runs has been varied between $N_{\text{sample}} = 1000$ for the largest and $N_{\text{sample}} = 5000$ for the smallest density, respectively, which insures that statistical fluctuations average out. During each run, the self and distinct density distributions are obtained with respect to each particle in the system, which obviously greatly enhances the efficiency of the method compared to working with a single self-particle.

The DMC results for the mean square displacement shown in Fig.~\ref{Fig_width1.0} were obtained from a suitable numerical integration of the simulation results for the density profile $\rho_s(r,t)$ of the self-particle. 

\section{Results} \label{SecResults}

  \begin{figure}
  \centering
  \includegraphics[width=9cm]{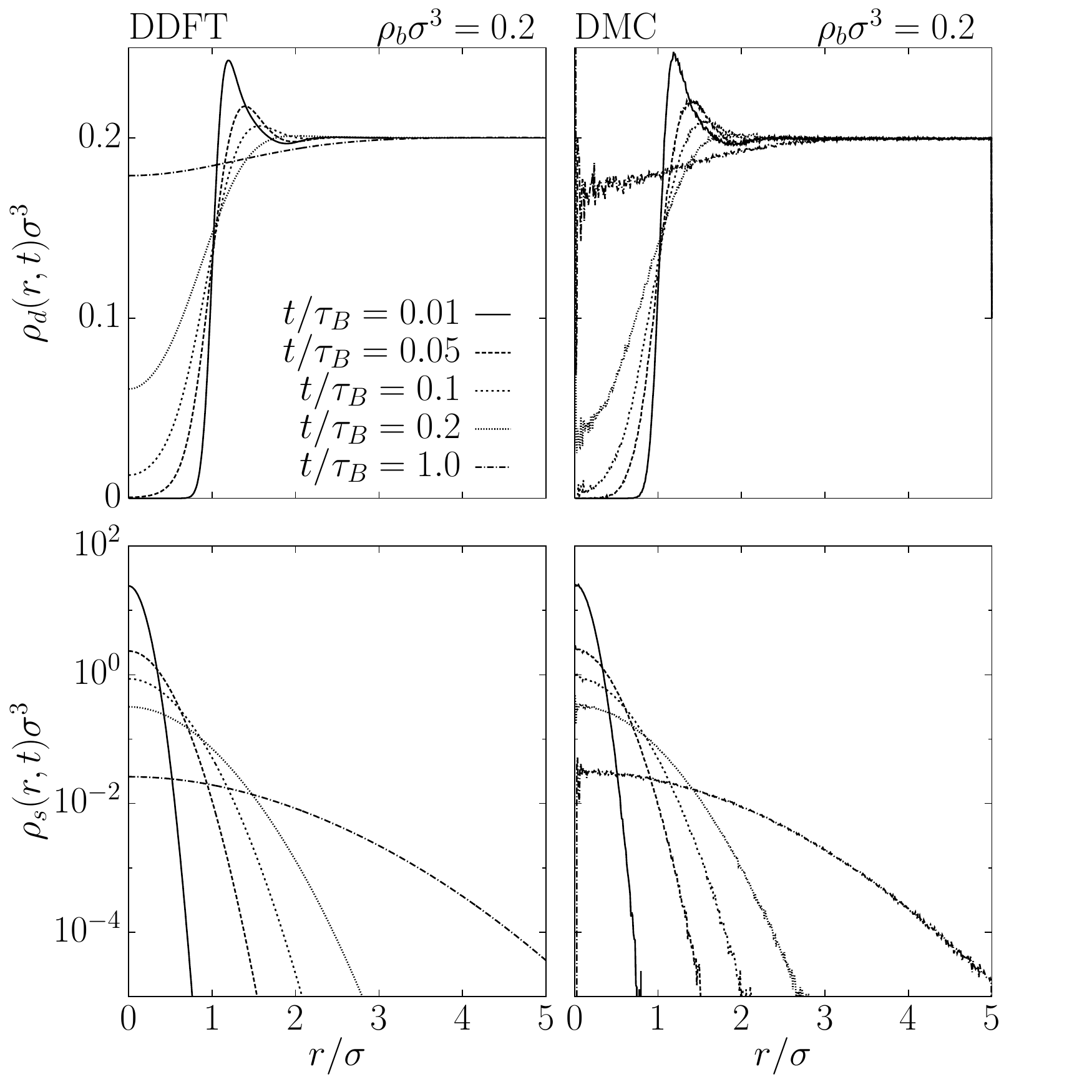}
  \caption{The self and distinct density profiles, $\rho_s(r,t)$ and $\rho_d(r, t)$, corresponding to the self and distinct parts of the van Hove function $\mathcal{G}(r, t)$ for a density  $\rho_b\sigma^3 = 0.2$. The plots on the left hand side show the results obtained from DDFT, on the right hand we display the DMC data. The lines correspond to times $t/\tau_B = 0.01, 0.05, 0.1, 0.2$ and $1$. Note that the ordinate axes of the self parts feature a logarithmic scale.}
  \label{Fig_density0.2}
  \end{figure}
  
    \begin{figure}
    \centering
  \includegraphics[width=9cm]{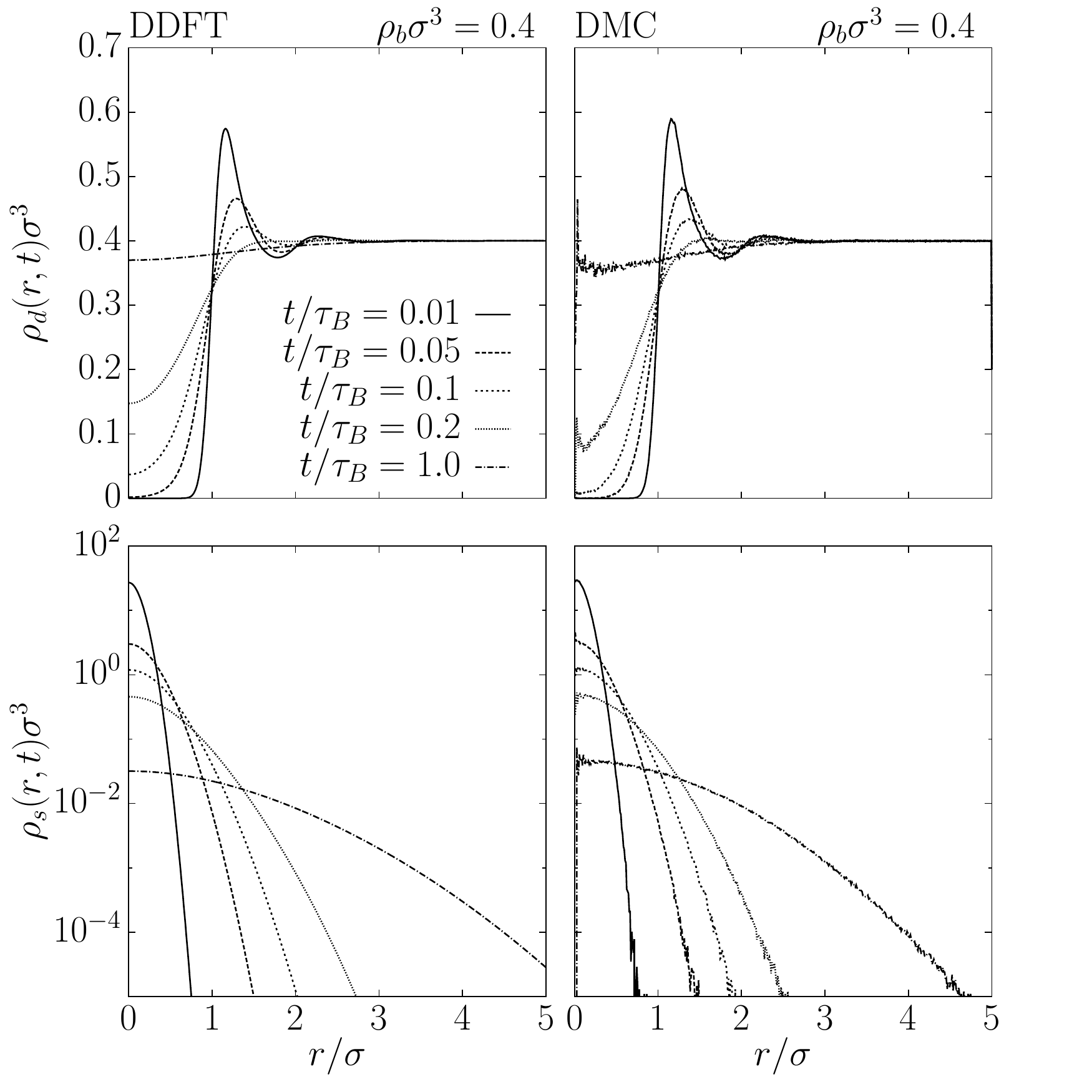} 
    \caption{The self and distinct density profiles, $\rho_s(r,t)$ and $\rho_d(r, t)$, for a density  $\rho_b\sigma^3 = 0.4$. The plots on the left hand side show the results obtained from DDFT, on the right hand we display the DMC data. The lines correspond to the same times as in Fig. \ref{Fig_density0.2}.}
    \label{Fig_density0.4}
    \end{figure}
      
 \begin{figure}[t!] 
 \centering
  \includegraphics[width=9cm]{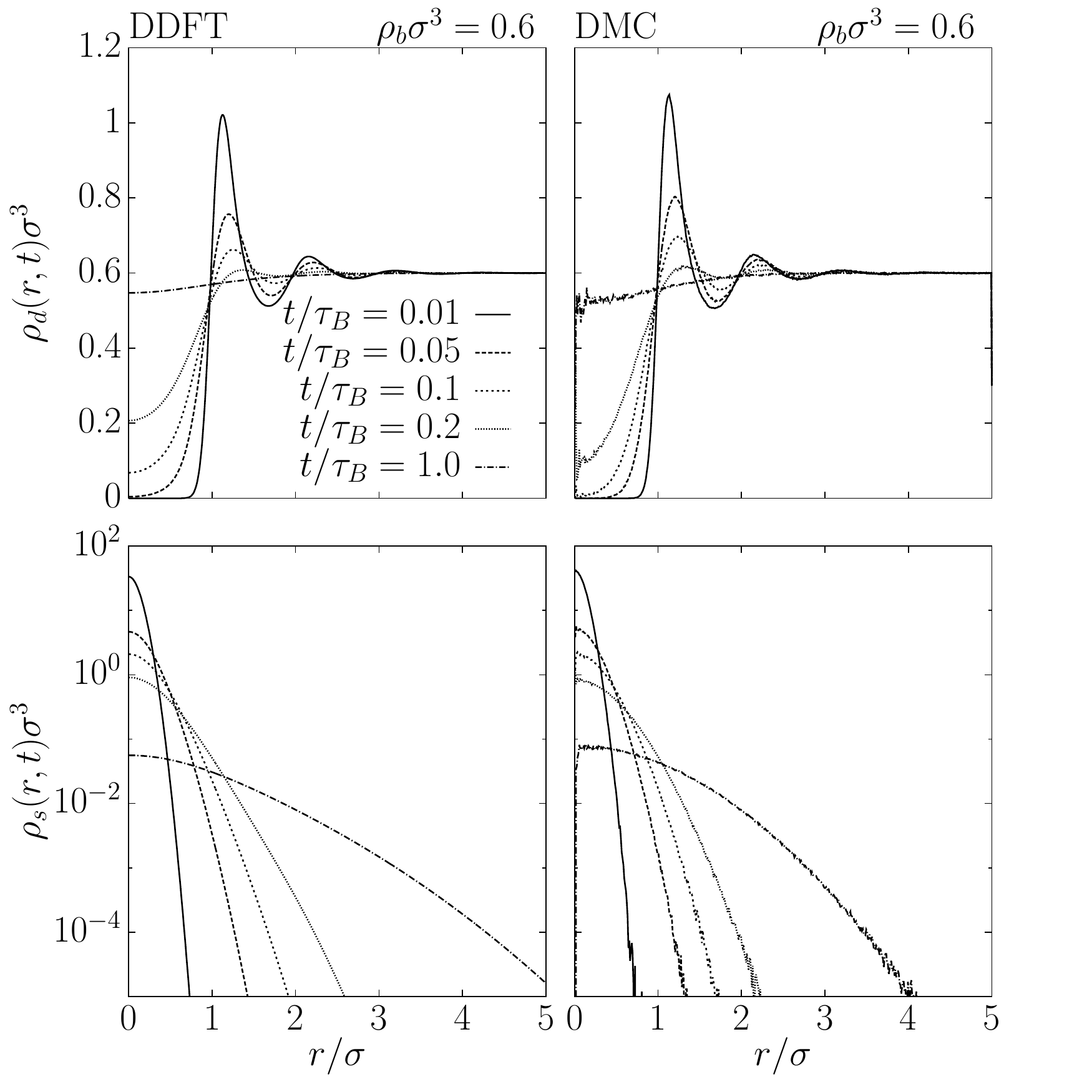}
 \caption{The self and distinct density profiles, $\rho_s(r,t)$ and $\rho_d(r, t)$, for a density  $\rho_b\sigma^3 = 0.6$. The plots on the left hand side show the results obtained from DDFT, on the right hand we display the DMC data. The lines correspond to the same times as in Fig. \ref{Fig_density0.2}.}
 \label{Fig_density0.6}
 \end{figure}
 
     \begin{figure}
     \centering
    \includegraphics[width=9cm]{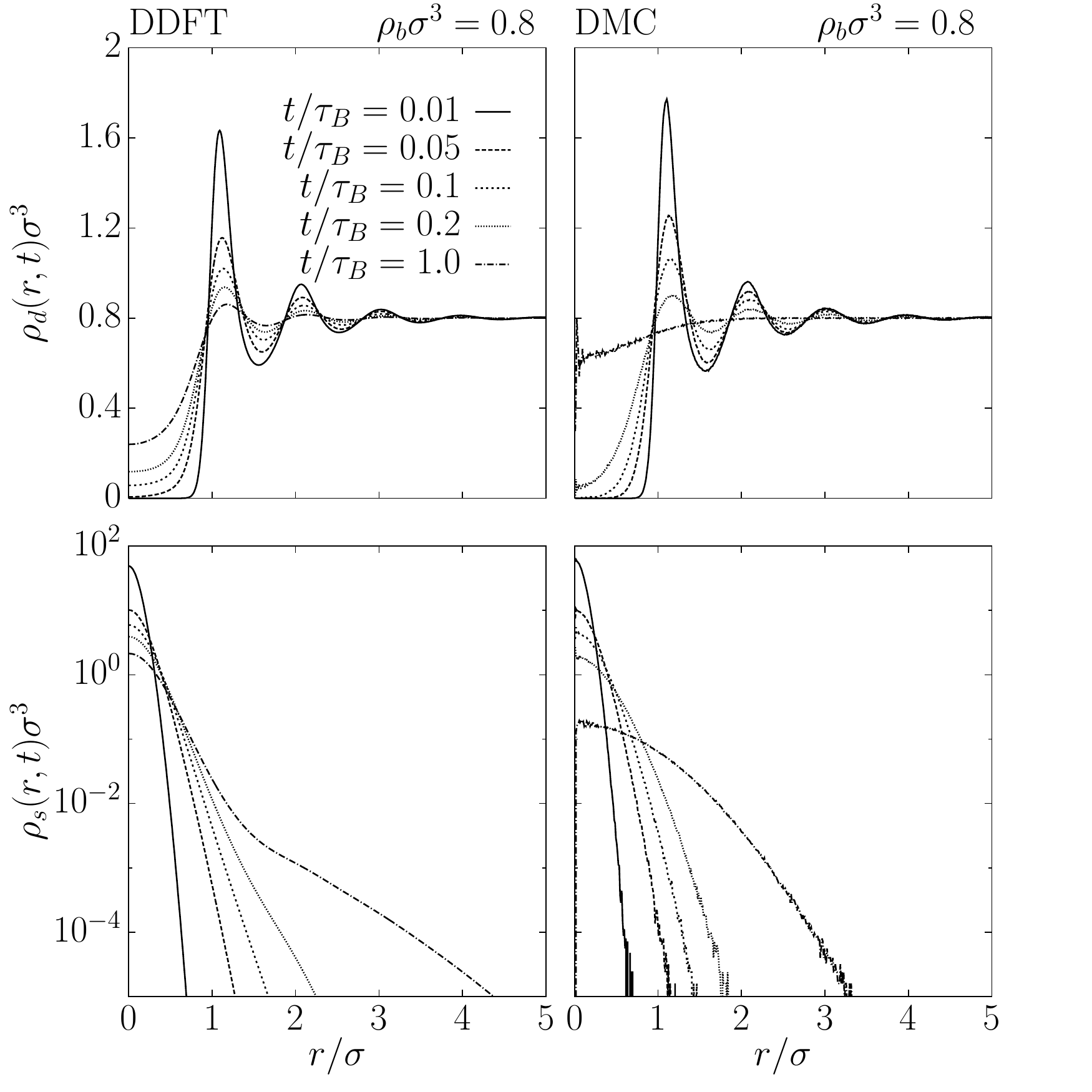}
    \caption{The self and distinct density profiles, $\rho_s(r,t)$ and $\rho_d(r, t)$, for a density  $\rho_b\sigma^3 = 0.8$. The plots on the left hand side show the results obtained from DDFT, on the  right hand we display the DMC data. The lines correspond to the same times as in Fig. \ref{Fig_density0.2}}
     \label{Fig_density0.8}
     \end{figure}

  \begin{figure}
  \centering
  \includegraphics[width=9cm]{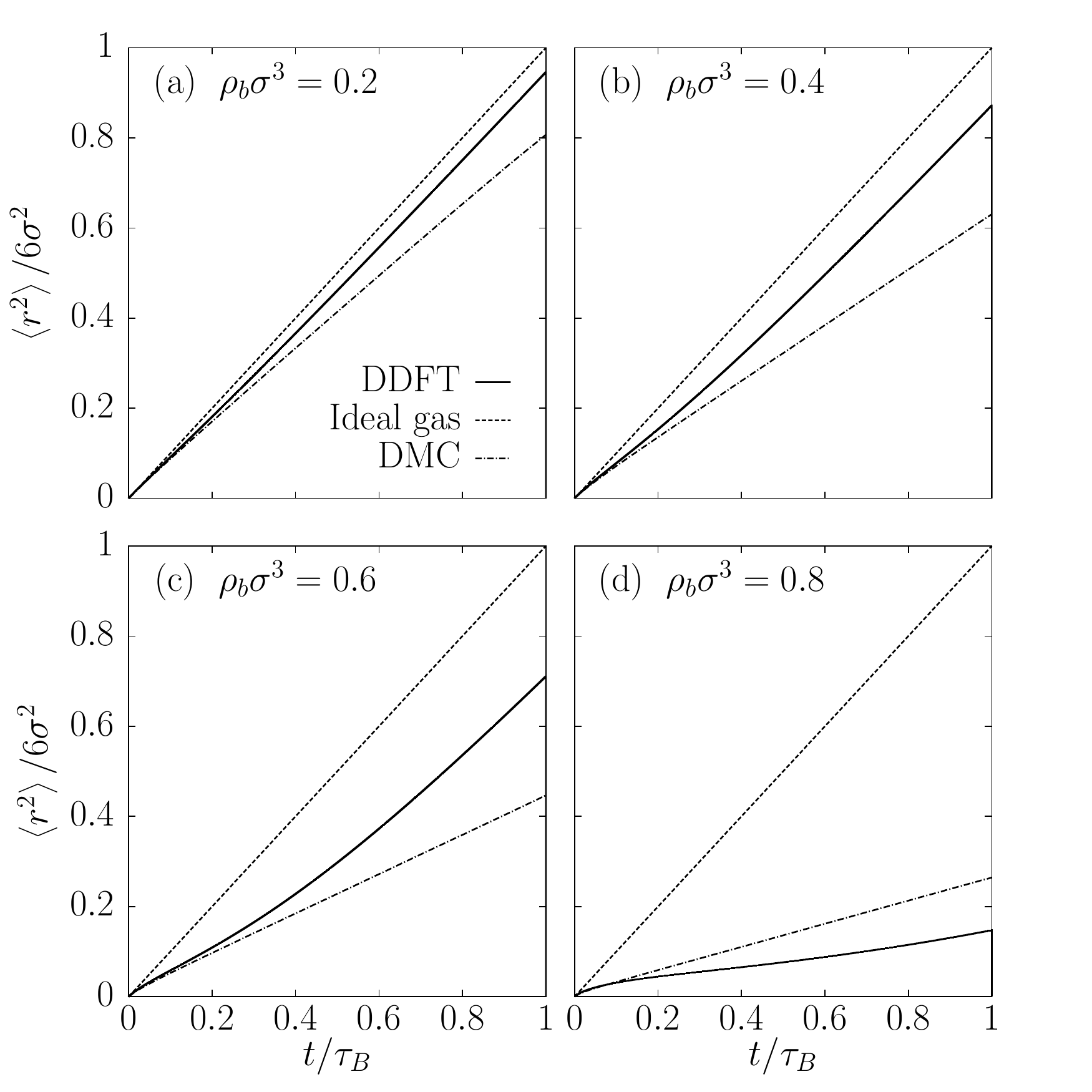}
  \caption{\footnotesize Mean square displacement $\braket{r^2}$ of the self part $\rho_s(r,t)$ as a function of $t/\tau_B$ for densities (a) $\rho_b\sigma^3 = 0.2$, (b) $\rho_b\sigma^3 = 0.4$, (c) $\rho_b\sigma^3 = 0.6$ and (d) $\rho_b\sigma^3 = 0.8$. The solid lines are DDFT results, the points represent data obtained by dynamic Monte-Carlo simulations. The data is normalized with respect to the ideal gas solution, Eq.~\eqref{EqMSQD}, thus the latter is a line with slope $1$.}
  \label{Fig_width1.0}
  \end{figure}

In Fig.~\ref{Fig_density0.2} we show the results for the self and distinct parts $\rho_s(r,t)$ and $\rho_d(r,t)$ of the van Hove function for a bulk density $\rho_b\sigma^3 = 0.2$. In the left column data obtained from dynamic test particle theory with the partial-linearization approach is shown, on the right hand side one can see the corresponding density profiles obtained from DMC. We performed calculations for times $t$ up to the Brownian time $\tau_B$. The ordinate axes of the self parts are shown on a logarithmic scale, thus a pure Gaussian distribution corresponds to a parabola. We find that there is very good qualitative as well as quantitative agreement between the DDFT results and the measured simulation data for all times, which is to be expected in the low-density limit considering that the present theory is exact for the ideal gas.

In Fig. \ref{Fig_width1.0} (a) the mean square displacement $\braket{r^2}$ of the self part as a function of $t/\tau_B$ up to $t = \tau_B$ is shown for a density $\rho_b\sigma^3 = 0.2$. Note that the results are normalized with respect to the ideal gas mean square displacement, thus the latter corresponds a line with slope 1. The simulations show that that $\braket{r^2}$ is reduced with respect to the ideal gas even for the rather low density of $\rho_b\sigma^3 = 0.2$. The DDFT result does follow the simulations rather well for short times while in the long-time limit the slope of the ideal gas is approached. Note that with a suitable $D_l < D$ the simulations are well described by Eq. \eqref{EqMSQD} for all times $t/\tau_B$ as can be seen from the linear dependence of $\braket{r^2}$ on $t$. 

In Fig.~\ref{Fig_density0.4} the obtained data for a bulk density $\rho_b\sigma^3 = 0.4$ are shown. As in Fig.~\ref{Fig_density0.2} we show results for $t$ up to the Brownian time $\tau_B$. The distinct density profiles obtained with the DDFT lose their structure somewhat too fast which is accompanied by a slightly wider self part than observed in the simulations. This trend has already been discussed in Sec~\ref{SubSecDDFTwithinFMT}. However, the agreement between simulations and DDFT is still very good. Moreover, our results in the low density regime $\rho_b\sigma^3 \lesssim 0.4$ deviate not much from those found by Hopkins {\it et al.} \cite{Schmidt2010}, which reflects the fact  that the RY approximation and a FMT based functional for the hard-sphere fluid coincide to lowest order in the density.

It can be seen from Fig.~\ref{Fig_width1.0} (b) that the mean square displacement $\braket{r^2}$ of the test particle as obtained from the DDFT is in good agreement with the simulations for short times while for long times the slope of $\braket{r^2}$ approaches that of the ideal gas, thus yielding a diffusion coefficient which is too large.
The convergence to the diffusivity of the ideal gas for large $t$ can be extracted analytically from the DDFT equations by studying the dynamics of a poorly localized particle in the bulk fluid. This amounts to prescribing a very broad Gaussian distribution for $\rho_s(r,t=0)$ while the background is initially given as $\rho_d(r,t=0) = \rho_b - \rho_s(r,t=0)$. Likewise, it can be shown analytically that for very short times $t \to 0$ the present DDFT yields the diffusivity of the ideal gas as well. Indeed, assuming that hydrodynamic interactions can be neglected, the initial stage of diffusion of hard spheres is that of a free particle \cite{BlPeMaDh92}. This can be understood from the short range of the hard core potential which causes hard spheres to not experience interactions (on average) on time scales significantly less than the time required to diffuse the mean nearest neighbor distance. The latter being on the order of $\sigma$ \cite{To95} we expect a crossover at a time less than $\tau_B$ from free diffusion to long-time diffusion which is slowed down by steric interactions. The simulations give a diffusivity near that of the ideal gas for $t<0.01\tau_B$ in agreement with the DDFT. However, while the agreement of the DDFT with simulations is still good in the range $0.01\tau_B<t<0.1\tau_B$, the subsequent asymptotic decay of the diffusion coefficient to the reduced long-time value is not captured by the DDFT which, as noted above, yields the diffusivity of the ideal gas for $t\to\infty$. We conclude that the structural information provided by the initial density profile $\rho_d(r, t=0)$ is required for the DDFT to yield a realistic time scale for the decay of $\rho_s(r, t)$. Once $\rho_d(r, t)$ becomes sufficiently close to the flat bulk profile, and hence structural information is lost, standard DDFT is bound to incorrectly yield the ideal gas diffusivity even in dense hard-sphere fluids.

In Fig. \ref{Fig_density0.6} we display the results for the self and distinct part for a intermediate density $\rho_b\sigma^3 = 0.6$ for times $t/\tau_B = 0.01, 0.05, 0.1$, $0.2$ and $1$. On the left hand side we show the results obtained by means of DDFT, on the right hand side one sees the corresponding DMC density profiles. We see that there is still good qualitatively agreement between DDFT and DMC for all times $t$. The distinct density profiles show a similar amount of structure in comparison with the DMC data, but we see that DDFT slightly underestimates the amount of structure in $\rho_d$ at early times. That is in contrast to the results found using the RY approximation, where at $\rho_b\sigma^3 = 0.6$ the functional starts to overestimate the amount of structure in the distinct part \cite{Schmidt2010}. The self part obtained within our DDFT is still well described by a Gaussian distribution. However, the ``speeding up'' effect of $\braket{r^2}(t)$ to the level of the ideal gas is now more pronounced than at lower densities and can be observed after relatively short times $t/\tau_B \lesssim 0.3$ (see Fig. \ref{Fig_width1.0} (c)). As already found at $\rho_b\sigma^3 = 0.4$ the slope of $\braket{r^2}(t)$ approaches unity at larger $t$ which corresponds to the diffusion constant of the ideal gas.

At a large density of $\rho_b\sigma^3 = 0.8$ (see Fig.~\ref{Fig_density0.8}) we find that DDFT yields reliable results for the self and distinct part until times $t/\tau_B \sim 0.2$. Compared to lower densities, the underestimation in the amount of structure of the distinct part in comparison to the DMC data intensifies. Furthermore, we see that the self part calculated from the DDFT is still well described by a Gaussian distribution up to times $t/\tau_B \sim 0.2$. For larger times, however, the self part shows a serious deviation from Gaussian shape that manifests itself in a ``fat tail'', see Fig.~\ref{Fig_density0.8}, corresponding to an exponential decay of $\rho_s(r,t)$. Interestingly, this behavior is indeed observed in soft matter systems, for instance in Brownian motion in supercooled liquids or close to jamming transitions, see Ref.~\onlinecite{WaKuBaGr12} for an overview of the subject matter. We observe that, as $t$ approaches $\tau_B$, the density profile of the distinct part loses its structure too slowly, which is in contrast to the behavior found at lower densities; this can also be observed from the behavior of the mean square displacement in Fig.~\ref{Fig_width1.0} (d), where the DDFT underestimates the simulation result. Interestingly, for times $t/\tau_B \gtrsim 2$ the mean square displacement shows a behavior $\sim t^2$ which is in contradiction with normal diffusion of the test particle. However, this behavior is required for the curve to reach a slope of one, which is indeed observed for times $t/\tau_B \gtrsim 5$ (not shown here), in agreement with the result from the analytical treatment mentioned above. The unrealistic slowdown of the diffusion, which is clearly seen from the subdiffusive behavior at intermediate times in Fig.~\ref{Fig_width1.0} (d) is strikingly reminiscent of the complete freezing of the dynamics which is observed by Hopkins {\em et al.} \cite{Schmidt2010} for a density $\rho_b\sigma^3 = 0.8$ for $t/\tau_B \gtrsim 0.1$ using the simple RY approximation for the hard-sphere free energy. In conclusion we tend to attribute the slowdown observed in the present work to inaccuracies of the FMT based free energy used in our DDFT, possibly caused by inconsistencies of the empirical partial-linearization approach, see Sec.~\ref{SubSecDDFTwithinFMT} for details. However, we would like to stress that while the slowdown at intermediate times is inaccurate the fact that there is qualitative agreement with simulations at $\rho_b\sigma^3 = 0.8$, meaning that our DDFT does not predict dynamic arrest, is a major step forward from previous, cruder implementations of DDFT. In light of these findings, the interpretation of the previously observed dynamical arrest in terms of a signature of glass-like behavior has to be reconsidered. We will dwell on this issue and related questions in the following final section.

\section{Summary and Conclusion} \label{SecConclusion}

In this work we studied the dynamic behavior of colloidal suspensions by means of the van Hove function which we calculate from a theoretical approach that is based on dynamic density functional theory (DDFT). The interaction between the colloids is assumed to be such that it can be mapped on an (effective) hard sphere model. In order to describe the free energy landscape of the non-uniform (equilibrium) system we employ fundamental measure theory (FMT) \cite{Rf89, Ro10}. More precisely, we use the very accurate White Bear II functional with the tensorial approach due to Tarazona \cite{Rf89, HaRo06, Tarazona2000}. To our knowledge this is the first study of the van Hove function of a hard-sphere colloidal system making use of an accurate FMT free energy functional. Within the framework of dynamic test particle theory \cite{Schmidt2007, Schmidt2010} we calculate the self and distinct parts of the van Hove function. In order to remove the interactions within the self part, which consists of only one particle, our first approach uses the analogy between our situation and the colloid-ideal polymer functional derived by Schmidt \textit{et al.} \cite{Schmidt2000}. It has been shown that such a colloid-ideal polymer functional can be derived by linearizing the excess free energy density with respect to the non-interacting component; in our case the self part of the van Hove function. As discussed in Sec.~\ref{SubSecDDFTwithinFMT}, mapping our system onto the colloid-polymer mixture scenario has a shortcoming. Due to the grand canonical character of the DFT we can only be sure that there is {\em one} self particle {\em on average} within the associated zero dimensional cavity. However, in our situation there is always {\em exactly one} test particle present.  It appears that losing the constraint $\eta_s + \eta_d < 1$ on the local packing fraction due to the linearization, results in density profiles of the distinct part which predict too high densities close to the test particle, what in turn leads to a rapid loss in the amount of structure, as can be seen in Fig.~\ref{Fig_fullVsPartLin}. Therefore, we introduced the partial-linearization approach, in which we calculate the direct correlation function of the distinct part $c^{(1)}_d(r,t)$ with the full, non-linearized functional, which guarantees to avoid violations of the local packing constraint.

We can conclude from comparison with data from our dynamic Monte Carlo (DMC) simulations that dynamic test particle theory combined with DDFT and FMT provides a reliable method for the calculation of the van Hove function of hard-sphere colloidal suspensions with densities up to $\rho_b\sigma^3 \lesssim 0.8$, i.e. packing fractions of around $40\%$. 
We observe that using FMT together with our empirical partial-linearization route yields a significant improvement in comparison to earlier results obtained with a simpler excess free energy model \cite{Schmidt2010}, in particular at intermediate and high densities. However, some deviations of our formulation of DDFT and the DMC simulations are still found.  We believe that both the slowdown effect within the diffusion of the self part and the overestimation of structure within the distinct part, which occur at intermediate times at a density $\rho_b\sigma^3 = 0.8$ (see Fig. \ref{Fig_width1.0} (d) and Fig. \ref{Fig_density0.8}), are likely artifacts of the empirical partial-linearization approach, caused by inconsistencies related to effectively using two different functionals. Importantly, in contrast to earlier work \cite{Schmidt2010} we no longer observe dynamic arrest at $\rho_b\sigma^3 = 0.8$. This freezing of the dynamics, which was interpreted as an indication of a glass transition, is therefore probably the signature of the overly simple free energy model for the hard-sphere mixture used in Ref.~\onlinecite{Schmidt2010}. However, dynamic arrest is observed in our model at $t\approx 0.2 \tau_B$ as we increase the density to $\rho_b\sigma^3 = 0.9$.

Moreover, as mentioned in Sec. \ref{SecResults}, the diffusivity obtained from the mean square displacement of the test particle approaches that of the ideal gas (see Fig. \ref{Fig_width1.0} (a) - (c)) in the long-time limit, in contrast to the simulation data which yield a long time self diffusion coefficient $D_l$ smaller than that of the ideal gas which decreases as the density is increased. This behavior causes an underestimation in the amount of structure of the distinct part and an overestimation of the width of the self part as can seen e.g.\ in Fig.~\ref{Fig_density0.6}. This shortcoming could in principle be fixed by using a mobility $\Gamma$ which is not a constant but rather allows the system to respond to local packing effects. One could for instance postulate that $\Gamma = \Gamma[\rho (\mathbf{r}, t)]$, thereby effectively making $\Gamma$ a function of space and time. We believe that this route has the potential of correcting the long-time diffusivity observed within standard DDFT.

We conclude that with the present work we have made an important step toward a comprehensive description of the dynamics of suspensions of colloidal hard spheres that would be of an accuracy equaling that of the FMT description of equilibrium hard-sphere systems. In order to fully obtain this goal, however, future work building on the framework introduced in this article will be necessary. We are confident that such improvements will be possible in the future, regarding in particular the search for a unique functional replacing the partial-linearization route and a more appropriate implementation of particle mobility, such that a comprehensive account of Brownian dynamics of colloidal particles within DDFT will be achieved.

\end{document}